\documentclass[twocolumn,pra,aps,amssymb,groupaddress,showpacs]{revtex4}
\usepackage{amsmath,amssymb,amsfonts}
\usepackage[usenames]{color}
\usepackage{graphicx,tikz,pgf,epsf}
\newcommand{\qed}{\hfill \rule{2mm}{2mm}}
\newcommand{\pf}{{\bf Proof: }}

\newtheorem{proposition}{Proposition}

\begin{document}
\title{Improved quantum test for linearity of a Boolean function}
\author{Kaushik Chakraborty and Subhamoy Maitra}
\affiliation{Indian Statistical Institute, 203 B T Road, Kolkata 700 108, India,\\
Email: kaushik.chakraborty9@gmail.com, subho@isical.ac.in\\
}
\begin{abstract}
Let a Boolean function be available as a black-box (oracle) and one likes to 
devise an algorithm to test whether it has certain property or it is 
$\epsilon$-far from having that property. The efficiency of the algorithm is 
judged by the number of calls to the oracle so that one can decide, with high 
probability, between these two alternatives. The best known quantum algorithm 
for testing whether a function is linear or $\epsilon$-far 
$(0 < \epsilon < \frac{1}{2})$ from linear functions requires 
$O(\epsilon^{-\frac{2}{3}})$ many calls [Hillery and Andersson, Physical Review
A 84, 062329 (2011)]. We show that this can be improved to 
$O(\epsilon^{-\frac{1}{2}})$ by using the Deutsch-Jozsa and the Grover 
Algorithms.
\end{abstract}
\maketitle

\section{Introduction}
Consider the model where a Boolean function is implemented inside a black box 
and we can obtain the output given an input in constant time. One such 
operation may be referred to as a query. Now we like to test several properties
of the Boolean function by exploiting such queries. Naturally, we like to test 
the properties with as less many queries as possible. In this paper we consider
algorithms to test whether a Boolean function is linear or it is $\epsilon$-far
from linear functions. The best known classical algorithm (BLR test~\cite{blr})
for testing this with good success probability requires $O(\frac{1}{\epsilon})$
query complexity. Naturally one should expect improvement in the quantum 
paradigm and in~\cite{hil1}, it has been shown that this can be tested with 
$O(\frac{1}{\epsilon^{\frac{2}{3}}})$ query complexity. In this paper, we
improve this further and show that this can be achieved with 
$O(\frac{1}{\epsilon^{\frac{1}{2}}})$ query complexity. 

\subsection{Basics of Boolean functions}
A Boolean function on $n$ variables may be viewed as a mapping from 
$V_n = \{0, 1\}^n$ into $\{0, 1\}$. The {\em truth table} of a Boolean function
$f(x_1, \ldots, x_n)$ is a binary string of length $2^n$, 
$f = [f(0, 0, \cdots, 0)$, $f(1, 0, \cdots, 0)$, $f(0, 1, \cdots, 0)$, $\ldots,
f(1, 1, \cdots, 1)]$. Let $\Omega_n$ be the set of all $n$-variable Boolean 
functions and it is easy to note that $|\Omega_n| = 2^{2^n}$.

The {\em Hamming weight} of a binary string $St$ is the number of 1's in $St$ 
denoted by $wt(St)$. An $n$-variable function $f$ is
said to be {\em balanced} if its truth table contains an equal
number of 0's and 1's, i.e., $wt(f) = 2^{n-1}$. Also, the {\em
Hamming distance} between equidimensional binary strings $St_1$ and
$St_2$ is defined by $d(St_1, St_2) = wt(St_1 \oplus St_2)$, where
$\oplus$ denotes the addition over $GF(2)$.

An $n$-variable Boolean function $f(x_1, \ldots, x_n)$ can be considered to be 
a multivariate polynomial over $GF(2)$. This polynomial can be expressed as  
$GF(2)$ sum of products representation of all distinct $k$-th order product 
terms
$(0 \leq k \leq n)$ of the variables. More precisely, $f(x_1, \ldots, x_n)$ 
can be written as $$a_0 \oplus \bigoplus_{1 \leq i \leq n} a_i x_i \oplus
\bigoplus_{1 \leq i < j \leq n} a_{ij} x_i x_j \oplus \ldots
\oplus a_{12\ldots n} x_1 x_2 \ldots x_n,$$ where the coefficients
$a_0, a_{ij}, \ldots, a_{12\ldots n} \in \{ 0, 1\}$. This representation of $f$
is called the {\em algebraic normal form} (ANF) of $f$. The number of variables
in the highest order product term with nonzero coefficient is called the 
{\em algebraic degree}, or simply the degree of $f$ and denoted by $deg(f)$.

Functions of degree at most one are called {\em affine} functions. An affine
function with constant term equal to zero is called a {\em linear} function.
The set of all $n$-variable affine functions is denoted by $A_n$. That is the 
set of affine functions contains all the linear functions and their complements.
 
Let ${x}=(x_1,\ldots,x_n)$ and ${a}=(a_1, \ldots, a_n)$ both belong to 
$\{0,1\}^n$ and the inner product 
$${a}\cdot{x} = a_1 x_1 \oplus \cdots \oplus a_n x_n.$$ 
A Boolean function $l(x)$ is called linear if it can be written as 
$l(x) = {a}\cdot{x}$ for some fixed $a$. Testing whether a Boolean 
function (given as an oracle) is linear or not is an important question 
in the field of computational complexity~\cite{BV93}. 

Let $f({x})$ be a Boolean function on $n$ variables. Then the {\em Walsh 
transform} of $f({x})$ is a real valued function over $\{0,1\}^n$ which is 
defined as 
$$W_f({\omega})=\sum_{{x}\in\{0,1\}^n}(-1)^{f({x})\oplus{\omega}\cdot{x}}.$$
One may also note the Parseval's relation in this case, which is
$$\sum_{{\omega}\in\{0,1\}^n} W^2_f(\omega) = 2^{2n}.$$
We also like to define the normalized Walsh transform as 
$$NW_f({\omega})=\frac{1}{2^n}\sum_{{x}\in\{0,1\}^n}(-1)^{f({x})\oplus{\omega}\cdot{x}}.$$
It is easy to check that $\sum_{{\omega}\in\{0,1\}^n} NW^2_f(\omega) = 1.$

The non-linearity (or non-affinity) of an $n$-variable function $f$ is 
$$nl(f) = min_{g \in A(n)} (d(f, g)),$$ i.e., the distance from the set of all 
$n$-variable affine functions. In terms of Walsh spectrum, the non-linearity of 
$f$ is given by 
$$nl(f)=2^{n-1} - \frac{1}{2}\max_{\omega \in \{0, 1\}^n}|W_f(\omega)|.$$

\subsection{Connection of Walsh spectrum with Deutsch-Jozsa algorithm}
Distinguishing constant and balanced Boolean functions with constant query 
complexity has been an important landmark in quantum computational 
framework that is well known as Deutsch-Jozsa algorithm~\cite{qDJ92}. 
Now let us discuss the relation between Deutsch-Jozsa algorithm and the 
Walsh Spectrum of a Boolean function, which is one of the important tools
in our work. It is known that given a classical circuit $f$, there is a 
quantum circuit of comparable efficiency which computes the transformation 
$U_f$ that takes input like $|x, y\rangle$ and produces output like 
$|x, y \oplus f(x)\rangle$. Let 
$|-\rangle = \frac{|0\rangle - |1\rangle}{\sqrt{2}}$. Then $U_f$ applied to the
state $|x\rangle|-\rangle$ will produce $(-1)^{f(x)} |x\rangle|-\rangle$. For 
brevity, we drop $|-\rangle$ and by abuse of notation, we denote that $U_f$ 
takes $|x\rangle$ to $(-1)^{f(x)} |x\rangle$.

Let $f$ be either constant or balanced and the corresponding quantum 
implementation $U_f$ is available. Deutsch-Jozsa~\cite{qDJ92} provided a 
quantum algorithm that can decide in constant time which one it is. Let us now 
describe another interpretation of Deutsch-Jozsa algorithm in terms of Walsh 
spectrum values~\cite{maitra}. We denote the operator for Deutsch-Jozsa 
algorithm as 
\begin{equation}
{\cal D}_f = H^{\otimes n} U_f H^{\otimes n}, \nonumber
\end{equation}
where the Boolean function $f$ is available as an oracle $U_f$. As in the case 
of $U_f$, for brevity, we abuse the notation and do not write the auxiliary 
qubit, i.e., $|-\rangle$ and the corresponding output in this case. 
Now one can observe that~\cite{maitra}
\begin{eqnarray}\nonumber
{\cal D}_f |0\rangle^{\otimes n}    &   =   &   \sum_{z \in \{0, 1\}^n}\sum_{x \in \{0, 1\}^n} \frac{(-1)^{x \cdot z \oplus f(x)}}{2^n}|z\rangle \\ \nonumber
                                    &   =   &   \sum_{z \in \{0, 1\}^n} \frac{W_f(z)}{2^n} |z\rangle \\ \nonumber
                                    &   =   &   \sum_{z \in \{0, 1\}^n} NW_f(z) |z\rangle. \nonumber
\end{eqnarray}
Note that the associated probability with a state $|z\rangle$ is 
$\frac{W_f^2(z)}{2^{2n}} = NW_f^2(z)$. 

\subsection{Background on linearity testing}
Let $l$ be a linear $n$-variable Boolean function, i.e., 
$l(x) = \omega \cdot x$, available in the form of an oracle and one likes
to get the $\omega$. For a linear function $l(x) = \omega \cdot x$,
$W_l(\omega) = 2^n$ and $W_l(z) = 0$, for $z \neq \omega$. Thus the observed
state of $n$ bits will clearly output $\omega$ itself 
(with probability $\frac{W_f^2(\omega)}{2^{2n}} = 1$). That is, the 
Deutsch-Jozsa algorithm solves this problem in constant time. In classical 
model, we need $O(n)$ time to find out the $\omega$. This difference and 
related results have been pointed out in~\cite{BV93}.

Now let us come to the question of testing linearity of a Boolean function.
This is a problem in the area of property testing and we refer 
to~\cite{bel,hil1,kauf} for further pointers in this specific area. There 
is huge literature in the area of property testing in general, e.g.,  
one may also refer to~\cite{childs} towards ideas in quantum property testing 
for bounded-degree graphs.

Given two 
$n$-variable Boolean functions $f$ and $g$, we define $f, g$ as $\epsilon$-far 
if $\frac{|\{x \in \{0, 1\}^n : f(x) \neq g(x)|\}}{2^n} = \frac{d(f, g)}{2^n} 
\geq \epsilon$, i.e., 
if the Hamming distance between the truth tables of $f$ and $g$ is at least 
$\epsilon 2^n$. Further, an $n$-variable Boolean function $f$ will be called 
$\epsilon$-far from a subset $S$ of $n$-variable Boolean functions if $f$ is 
$\epsilon$-far from all the functions $g \in S$. 

The probabilistic classical test for linearity is well known as the BLR 
test~\cite{blr} that exploits the condition $l(x \oplus y) = l(x) \oplus l(y)$ 
for a linear function $l$, where $a_0 = 0$. However, if $a_0 = 1$ for an 
affine function $\ell$, then we have the condition 
$\ell(x \oplus y) = 1 \oplus \ell(x) \oplus \ell(y)$. One may note that
one can easily decide whether $a_0 = 0$ or $1$ by checking the output of
the function at the all-zero, i.e., $(0, 0, \ldots, 0)$ input. Thus the 
probabilistic classical algorithm for testing whether an $n$-variable Boolean 
function $f$ is affine or not works as follows.

\ \\
{\bf Algorithm 1.}
\begin{enumerate}
\item Let $a_0 = f(0, 0, \ldots, 0)$.
\item For $t$ many times 
\begin{enumerate}
\item Randomly choose distinct $x, y \in \{0, 1\}^n$.
\item Check the condition $f(x \oplus y) = a_0 \oplus f(x) \oplus f(y)$. 
\item If the condition is not satisfied, report that $f$ is not
affine and terminate.
\end{enumerate}
\item Report that the function is affine and terminate.
\end{enumerate}
It is well known that if the algorithm reports that $f$ is non-affine,
then it is non-affine with probability $1$, but if it reports that $f$ 
is affine, then it succeeds with some probability depending on the number of
iterations $t$. A simple analysis shows that if one needs to decide whether
a function is $\epsilon$-far from the set of affine functions, then the 
probability of success is greater than or equal to $\frac{2}{3}$ (or any 
constant $c$, such that $\frac{1}{2} < c < 1$) where $t$ is 
$O(\frac{1}{\epsilon})$. However, the detailed analysis of this probability
of success is quite involved and one may refer to~\cite{bel,kauf} in this 
direction. 

Consider that an $n$-variable function $f$ is $\epsilon$-far from $A_n$, the set
of all $n$-variable affine functions. That 
means, $nl(f) \geq \epsilon 2^n$. Using Parseval's result, it is easy to 
note that $nl(f) \leq 2^{n-1} - 2^{\frac{n}{2}-1}$. The upper bound can be 
achieved for functions on even number of variables which are known
as bent functions. However, the problem is yet to be settled for the cases
on odd number of variables. This tells a function on $n$ variables can
be $\epsilon$-far from the set of affine functions where 
$0 \leq \epsilon \leq \frac{1}{2} - \frac{1}{2^{\frac{n}{2}+1}}$. 
For details of combinatorial, cryptographic and coding theoretic results
related to Boolean functions, one may see~\cite{kavut} and the references
therein. In general, as $\frac{1}{2^{\frac{n}{2}+1}}$ tends to 0 for 
large $n$, we will consider $0 < \epsilon < \frac{1}{2}$ throughout this 
document.

Towards solving many computational problems, quantum algorithms provide 
improved query complexity and in that line a quantum algorithm is described 
in~\cite{hil1}, that achieves an improved 
query complexity $O(\frac{1}{\epsilon^{\frac{2}{3}}})$. We noted that this
problem is related to different variants of satisfiability 
problem~\cite{bel,kauf} and thus it may be natural to obtain a quadratic 
speed-up over the classical paradigm using Grover algorithm~\cite{qGR96}. 
We find that this is indeed true and present a probabilistic quantum algorithm 
that works in $O(\frac{1}{\epsilon^{\frac{1}{2}}})$ query complexity. 

\section{Our proposal}
\label{our}
We first present our basic idea using Deutsch-Jozsa~\cite{qDJ92} algorithm.

\ \\
{\bf Algorithm 2.}
\begin{enumerate}
\item Let $|\Psi\rangle = {\cal D}_f(|0\rangle^{\otimes n})$.
\item Measure $|\Psi\rangle$ in computational basis and let the measured
state be $a^{(0)}$ (an $n$-bit pattern).
\item For $t$ many times ($i = 1$ to $t$) 
\begin{enumerate}
\item Let $|\Psi\rangle = {\cal D}_f(|0\rangle^{\otimes n})$.
\item Measure $|\Psi\rangle$ in computational basis and let the measured
state be $a^{(i)}$ (an $n$-bit pattern).
\item If $a^{(i)} \neq a^{(0)}$, report that the function is not
affine and terminate.
\end{enumerate}
\item Report that the function is affine and terminate.
\end{enumerate}

If Algorithm 2 reports that a function is not affine, then it reports this
correctly. However, if it reports that the function is affine, that may
or may not be correct. If the function is affine, then Algorithm 2 reports it
correctly. However, there may be cases where the function is not affine, but 
still the algorithm reports it as an affine function. Consider that the 
function is $\epsilon$-far from $A_n$. Then one can check that 
$|NW_f(\omega)| \leq 1-2\epsilon$ for any $\omega \in \{0, 1\}^n$.
Thus, $|NW_f(a^{(0)})| \leq 1-2\epsilon$ too. To wrongly report the function
is affine, Algorithm 2 must report $a^{(i)} = a^{(0)}$ for all $i = 1$ to $t$. 
This happens with
probability $\leq (1-2\epsilon)^t$. Thus, it is easy to note that 
with $O(\frac{1}{\epsilon})$ many iterations, we can reduce the error
probability below $\frac{1}{3}$, i.e., the success probability will be 
greater than or equal to $\frac{2}{3}$. In terms of query complexity, 
this is same as the case for the classical BLR~\cite{blr} test. We will
now improve this algorithm towards better query complexity.

\subsection{Use of Grover Algorithm for further improvement}
Consider that a function is $\epsilon$-far from $A_n$.
In line of Grover algorithm~\cite{qGR96}, we will try to reduce the amplitude
corresponding to the state $|a^{(0)}\rangle$ and increase the amplitude
of the other states so that we can quickly obtain an $a^{(i)}$ after 
measurement, which is not equal to $a^{(0)}$.

Our idea is as follows. Instead of equal superposition 
$H^{\otimes n}(|0\rangle^{\otimes n}) = 
\frac{1}{2^{\frac{n}{2}}} \sum_{z \in \{0, 1\}^n} |z\rangle$ in Grover 
algorithm, we will use the state of the form 
$|\Psi\rangle = {\cal D}_f (|0\rangle^{\otimes n}) 
= \sum_{z \in \{0, 1\}^n} \frac{W_f(z)}{2^n} |z\rangle
= \sum_{z \in \{0, 1\}^n} NW_f(z) |z\rangle$.

Further, towards inverting the phase, we will use another $n$-variable Boolean 
function $g(x)$, different from $f(x)$, where $g(x) = 0$, when $x = a^{(0)}$, 
and $g(x) = 1$, otherwise. Based on $g(x)$, we implement the inversion 
operator as ${\cal O}_g$, that inverts the phase of the states $|x\rangle$ 
where $\{x \in \{0, 1\}^n : x \neq a^{(0)}\}$. Note that one can efficiently
implement $g(x)$ in classical domain with $O(n)$ many gates and thus we can
also implement ${\cal O}_g$ efficiently. Finally, we consider the operator 
$$G_t = \left[(2 |\Psi \rangle \langle \Psi| - I) {\cal O}_g \right]^t$$ on 
$|\Psi\rangle$ to get $|\Psi_t\rangle$. 

Consider the $n$-qubit state 
$|\Psi\rangle = \sum_{s \in S} u_s |s\rangle + \sum_{s \in \{0, 1\}^n 
\setminus S} v_s |s\rangle$, where $u_s, v_s$ are real and 
$\sum_{s \in S} u^2_s + \sum_{s \in \{0, 1\}^n \setminus S} v^2_s = 1$. 
For brevity, let us represent 
$|\Psi\rangle = \sum_{s \in S} u_s |s\rangle + 
\sum_{s \in \{0, 1\}^n \setminus S} v_s |s\rangle = 
u|X\rangle + v|Y\rangle$. That is, $u^2 = \sum_{s \in S} u^2_s$ 
and $v^2 = \sum_{s \in \{0, 1\}^n \setminus S} v^2_s$. In this general
framework, consider that $g(x) = 1$, when $x \in S$ and $g(x) = 0$ otherwise.
Now we have the following technical result.
\begin{proposition}
\label{prop1}
Let $u = \sin \theta$, $v = \cos \theta$. 
The application of $[(2|\Psi\rangle\langle\Psi| - I){\cal O}_g]^{t}$ operator 
on $|\Psi\rangle$ produces $|\Psi_t\rangle$, in which the 
amplitude of $|X\rangle$ is $\sin (2t+1)\theta$.
\end{proposition}
\pf For $t=1$, one can check that 
$|\Psi_{1}\rangle = [(2|\Psi\rangle\langle\Psi| - I) {\cal O}_g]|\Psi\rangle =
[(2|\Psi\rangle\langle\Psi|){\cal O}_g]|\Psi\rangle -{\cal O}_g|\Psi\rangle$. 
Now substituting the values of $u, v$, we get that 
$|\Psi_1\rangle = \sin 3\theta|X\rangle + \cos 3\theta |Y\rangle$.

Now we will use induction. Let the application of 
$[(2|\Psi\rangle\langle\Psi| - I){\cal O}_g]^{t}$ operator on $|\Psi\rangle$ 
updates the amplitude of $|X\rangle$ as $\sin (2t\theta + \theta)$,
for $t = k$. From the assumption we have 
$[(2|\Psi\rangle\langle\Psi| - I){\cal O}_g]^{k}|\Psi\rangle = 
\sin (\theta + 2k \theta)|X\rangle + \cos (\theta + 2k\theta)|Y\rangle$. 
Now, for $t=k+1$, it can be checked that 
$[(2|\Psi\rangle\langle\Psi| - I){\cal O}_g]^{(k+1)}|\Psi\rangle = 
\sin (\theta + 2(k+1) \theta)|X\rangle + \cos (\theta + 2(k+1)\theta)|Y\rangle$.
Thus, the proof. \qed

In our case, $S = \{0, 1\}^n \setminus \{a^{(0)}\}$.   
Let us now present our improved algorithm.

\ \\
{\bf Algorithm 3.}
\begin{enumerate}
\item Let $|\Psi\rangle = {\cal D}_f(|0\rangle^{\otimes n})$.
\item Measure $|\Psi\rangle$ in computational basis and let the measured
state be $a^{(0)}$ (an $n$-bit pattern).
\item Consider a Boolean function $g$ such that 
$g(x) = 0$, when $x = a^{(0)}$, 
and $g(x) = 1$, otherwise. 
\item Obtain $|\Psi_t\rangle = 
[(2|\Psi\rangle\langle\Psi| - I){\cal O}_g]^{t} (|\Psi\rangle)$.
(Note that $t$ is the significant complexity parameter in this algorithm.)
\item Measure $|\Psi_t\rangle$ in computational basis and let the measured
state be $a^{(t)}$ (an $n$-bit pattern).
\item If $a^{(t)} \neq a^{(0)}$, report that the function is not
affine and terminate.
\item Report that the function is affine and terminate.
\end{enumerate}

In this case, we use the state $|\Psi_t\rangle$ for measurement in 
computational basis. Consider that after the Deutsch-Jozsa algorithm we obtain 
an $n$-qubit
state $|\Psi_0\rangle$ (before the measurement) and observed $a^{(0)}$ after
measurement. In case the function in consideration is indeed affine, i.e., of 
the form $a_0 \oplus a^{(0)}\cdot x$ then $\sin \theta = 0$. Hence, 
amplitude of $|X\rangle$ (the quantum state which is the superposition of all
states except $|a^{(0)}\rangle$), after $t$ many iterations will remain as
$\sin (2t+1)\theta = 0$. Hence the measurement of the state $|\Psi_t\rangle$ 
in computational basis will provide $a^{(0)}$ again. In case $f$ is not 
affine, we have $\sin \theta > 0$. Thus, with proper choice of $t$, it is 
possible to obtain $\sin^2 (2t+1)\theta \geq \frac{2}{3}$ and hence the 
measurement of the state $|\Psi_t\rangle$ will provide $a^{(t)} \neq a^{(0)}$ 
with probability $\geq \frac{2}{3}$. 

Now the final point left is to show that $t$ is $O(\sqrt{\frac{1}{\epsilon}})$.
As we considered, let $|\Psi\rangle =  u|X\rangle + v|Y\rangle$. 
Here $|Y\rangle = |a^{(0)}\rangle$, i.e., $v \leq 1 - 2\epsilon$.
Thus, $u \geq \sqrt{1 - (1-2\epsilon)^2} = \sqrt{4\epsilon - 4 \epsilon^2}
\geq \sqrt{2\epsilon}$ as $\epsilon < \frac{1}{2}$.
We take $u = \sin \theta$, $v = \cos \theta$. That is 
$\sin \theta \geq \sqrt{2\epsilon}$. Considering $\theta$ small, we can write 
$\sin \theta \approx \theta$ and we want $t$ such that 
$(2t+1)\theta \approx \frac{\pi}{2}$. In this case, $\sin (2t+1)\theta$ becomes
close to $1$ (it is enough to get $\sin^2 (2t+1)\theta \geq \frac{2}{3}$ or 
some constant greater than $\frac{1}{2}$). Thus, it is immediate to note that 
$t$ should be $O(\sqrt{\frac{1}{\epsilon}})$. This completes the analysis for 
the query complexity of Algorithm 3. 

\section{Conclusion and Open Problems}
\label{conclusion}
In this paper we present a quantum algorithm to test whether a function is 
affine or it is $\epsilon$-far $(0 < \epsilon < \frac{1}{2})$ from the set of 
affine functions. While the best known classical algorithm~\cite{blr} requires 
$O(\frac{1}{\epsilon})$ query complexity and the existing quantum 
algorithm~\cite{hil1}
takes $O(\frac{1}{\epsilon^\frac{2}{3}})$, the query complexity of our
proposal is $O(\frac{1}{\epsilon^\frac{1}{2}})$. 

One important issue is how the complexity is related to $n$, the number of input
variables to the Boolean function in question. As we have discussed earier,
while testing for whether a function is $\epsilon$-far from the set of
$n$-variable affine functions $A_n$, we have 
$0 \leq \epsilon \leq \frac{1}{2} - \frac{1}{2^{\frac{n}{2}+1}}$. 
If a function is at a constant distance $\delta$ from $A_n$, then
$\epsilon = \frac{\delta}{2^n}$ and thus the Algorithm 3 will require
$O(2^{\frac{n}{2}})$ time complexity. If $\delta = \frac{2^n}{\zeta(n)}$, 
then the algorithm will require order of $\sqrt{\zeta(n)}$ time. That is,
if $\zeta(n)$ is polynomial in $n$, then we have a quantum probabilistic 
polynomial-time algorithm here. The algorithm will require constant time 
for highly nonlinear functions where $\delta$ is $O(2^n)$, i.e., 
when $\epsilon$ is constant.

Informally speaking, 
it is natural from the optimality results~\cite{opt} of the Grover algorithm, 
that lesser quantum query complexity than what we propose here for linearity 
testing may not be achievable. It is interesting to explore how this kind of
technique using Walsh spectrum of Boolean functions, associated with 
the Deutsch-Jozsa and the Grover Algorithms, can be exploited for testing 
some other properties of Boolean functions.

\end{document}